# Quantum simulated annealing

Steve Huntsman
Institute for Defense Analyses
*shuntsma@ida.org*
December 2000; revised July 2001

*Abstract.* The question of whether or not quantum computers can efficiently solve NP-complete problems is open, although many indications are negative. However, many of these problems are natural candidates for solution on quantum computers. Here we outline a thermodynamical formalism for the traveling salesman problem which allows for its polytime solution on a quantum computer with probability arbitrarily close to unity, with sufficient energy resources and subject to a weak nondegeneracy constraint on the distances. Applications to other problems are also discussed.

The question of whether or not quantum computers can efficiently solve NP-complete problems is at present unresolved: the prospects for general solutions of NP-complete problems are dim [BV], [BBB], [Ce2]. However, many instances of such problems—possibly including those which can be characterized as the hardest instances [KS]—are natural candidates for solution on quantum computers. [HP], [FG] Here we outline a thermodynamical formalism for the traveling salesman problem which allows for its polytime solution on a quantum computer almost surely, given sufficient energy resources, and also subject to a weak constraint on the distances.

Consider the traveling salesman problem (TSP) for $n$ cities. Now corresponding to each Hamiltonian path or tour of the complete graph (i.e., a path visiting every city only once) is a unique permutation $\sigma$, which in turn can be regarded as a series of $n$ edges encoded as pairs of numbers in such a way that the end of one edge is the beginning of its successor: therefore, since we can wlog ignore the last vertex, $n-1$ of the $n(n-1)/2$ edges are represented in a given permutation. By ignoring the last vertex we can map the TSP onto a problem on a tree from which $n$ edges stem from the root, $n-1$ edges from each of these, and so forth.

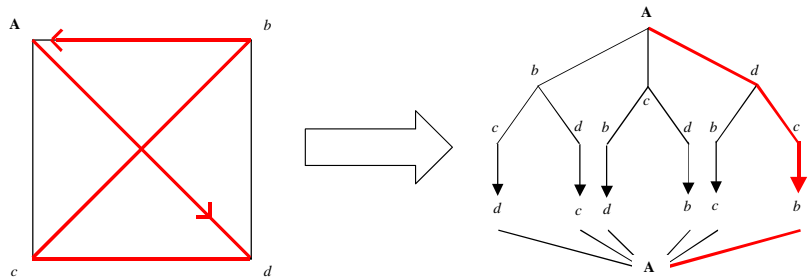

Fig. Turning a complete graph into a tree.

It is evident that by using controlled quantum logic operations we can construct a quantum superposition encoding the set of all tours on a complete graph. (A construction which is similar in spirit is in [Ru].) Such a quantum state could be (for instance) a superposition of bitstrings of the form

$$1_{(1)}1_{(2)}\ldots 0_{\sigma(1)}\ldots 1_{(n)}\ 1_{(n+1)}1_{(n+2)}\ldots 0_{\sigma(2)}\ldots 1_{(2n)}\ \ldots\ldots 1_{(n(n-1))}1_{(n(n-1)+1)}\ldots 0_{\sigma(n)}\ldots 1_{(n(n))}$$

(so the zero at the $\sigma(j)$-labeled bit acts as a place marker for the $j$th city). Alternatively, we might build on a device of Abrams and Lloyd [AL], who showed how to efficiently realize a superposition of the form

$$\sum_{\sigma \in S_n} |\sigma(1,...,n)\rangle.$$

The idea now is to expand such a state encoding the tours into a state $|S\rangle$ which will allow us to apply gates weighing individual edges in parallel according to their component edge distances. (How we weigh the edges is crucial, since to obtain a minimal path requires us to isolate one of the $n!$ possible paths.) A method for obtaining such a suitable $|S\rangle$ can proceed either by invoking the tree structure of the TSP to directly encode the paths in a quadratic number of qubits (an analogue of this corresponding to a binary tree which allows edges to be addressed individually is relatively straightforward; this corresponds to simulating a tournament on a quantum computer [Hu]) or from the superposition of permutations referred to above. In either case this can be accomplished efficiently.

Consider the transformations

$$U_{jk} \equiv \begin{pmatrix} \sqrt{q_{jk}} & \sqrt{1-q_{jk}} \\ -\sqrt{1-q_{jk}} & \sqrt{q_{jk}} \end{pmatrix}$$

with $-\log_\alpha q_{jk} \equiv d_{jk}$, the normalized distance between cities $j$ and $k$ (so that $\max(d_{jk}) = 1$ and $1/\alpha \leq q_{jk} < 1$). These can be efficiently implemented [Ki]. Now

$$\Pr(|a\rangle) \equiv |\langle 0|U_{q_{jk}}|a\rangle|^2 = \begin{cases} q_{jk} & \text{if } |a\rangle = |0\rangle \\ 1-q_{jk} & \text{if } |a\rangle = |1\rangle \end{cases}$$

and so we can use these gates to bias a solution to the TSP given suitably encoded states (the need for which arises from the requirement that we apply the appropriate gates conditioned on what a given state encodes).

The $U$-operators do not preserve the bitstrings above; however, these bitstrings all share the common feature of having exactly $n$ zeros. This can be exploited by following the controlled application of $U$-operators by a projection onto the subspace with exactly $n$ zeros before a measurement. This destructive projection could be (we use an example with photons for simplicity) realized as a polarization filter applied prior to a measurement. Therefore the implicit cost in this scheme (and others like it) is energy: it turns out (see below) that if $\beta \equiv \log \alpha$ is the effective temperature for an annealing schedule, then we will require on the order of $\alpha^D$ photons (here $D$ is a distance scale for the TSP given below) for a successful realization of the algorithm. This generalizes an earlier result of [Ce2].

So by constructing appropriate initial states and applying the probability gates for each of the legs $\sigma(j)\sigma(j+1)$ (i.e., applying a gate iff the appropriate bit is zero), a representation of the entire space of candidate solutions which also encodes information about edge distances can be generated with a quadratic number of qubits and a polynomial number of gates. Measurements can be performed and the whole procedure repeated as many times as necessary or feasible. We now proceed to show what this can entail.

The object of the TSP is to find the permutation minimizing the total distance, given as the sum

$$D_\sigma \equiv \sum_{j=1}^{n} d_{\sigma(j)\sigma(j+1)}$$

which we can, in turn, express as

$$-\sum_{j=1}^{n} \log_\alpha q_{\sigma(j)\sigma(j+1)} = -\log_\alpha \prod_{j-1}^{n} q_{\sigma(j)\sigma(j+1)}.$$

So by maximizing the product of the "probabilities" (which is equivalent to maximizing the logarithm of their product) we can bias the solution of the TSP. This will also bias near-solutions.

As an illustration consider the four-city TSP with cities marked by the indices 1, 2, 3, 4 (wlog we will start our tours at 1) with distance parameters (using $\alpha = e$)

$$\begin{aligned} d_{12} &= .7, \quad q_{12} = .4966 \\ d_{13} &= .5, \quad q_{13} = .6065 \\ d_{14} &= 1, \quad q_{14} = .3679 \end{aligned}$$

$$d_{23} = .8, \quad q_{23} = .4493$$
$$d_{24} = .6, \quad q_{24} = .5488$$
$$d_{34} = .9, \quad q_{34} = .4066$$

which yields the following $\Pi q$ terms:

$$\sigma = (1)234(1) \Rightarrow \Pi q_{\sigma(j)\sigma(j+1)} = \exp(-D_\sigma) = .0334, D_\sigma = 3.4$$
$$\sigma = (1)243(1) \Rightarrow \Pi q_{\sigma(j)\sigma(j+1)} = \exp(-D_\sigma) = .0672, D_\sigma = 2.7$$
$$\sigma = (1)324(1) \Rightarrow \Pi q_{\sigma(j)\sigma(j+1)} = \exp(-D_\sigma) = .0550, D_\sigma = 2.9$$
$$\sigma = (1)342(1) \Rightarrow \Pi q_{\sigma(j)\sigma(j+1)} = \exp(-D_\sigma) = .0672, D_\sigma = 2.7$$
$$\sigma = (1)423(1) \Rightarrow \Pi q_{\sigma(j)\sigma(j+1)} = \exp(-D_\sigma) = .0550, D_\sigma = 2.9$$
$$\sigma = (1)432(1) \Rightarrow \Pi q_{\sigma(j)\sigma(j+1)} = \exp(-D_\sigma) = .0334, D_\sigma = 3.4$$

and these sum to $(Z =) .3112$, so the actual probabilities are respectively .1073, .2159, .1767, .2159, .1767 and .1073. The effect of the biasing is clear: a solution (there are two equivalent solutions for the symmetric TSP) to this instance of the TSP will be measured 43 percent of the time (i.e., $2 \cdot .0672/.3112 = 2 \cdot .2159 = .43$ is the probability of choosing a solution; without a bias, this probability is .33).

The question now becomes one of how much of a bias is required to observe the exact solution to the TSP in expected polytime. Let $\tau$ denote a permutation corresponding to a solution of the TSP; the naïve probability of observing the state corresponding to it will be $\Pr(\tau) = \Pi_j q_{\tau(j)\tau(j+1)}/\Sigma_\sigma \Pi_j q_{\sigma(j)\sigma(j+1)}$. The denominator arises from the formal partition function

$$\sum_{\sigma \in S_n} \prod_{j=1}^{n} q_{\sigma(j)\sigma(j+1)} = \sum_{\sigma \in S_n} \alpha^{-D_\sigma} = \sum_{\sigma \in S_n} e^{-D_\sigma \beta} \equiv Z(\beta)$$

where $\beta$ (which corresponds to a $1/k_B T$ or inverse temperature term in a partition function) $\equiv \log \alpha$. It is clear that without a significant bias the probability of measuring the solution will still in general be too low to be possible in polytime. Indeed we require that $\Pr(\tau) = O(n^{-k})$ for $k << n$ (rather than $\Pr(\tau) = O(1/n!)$) for an efficient ($O(n^k)$, rather than $O(n!)$) solution to the TSP.

This criterion for expected polytime solution (CP) holds if

$$\alpha^{-D_\tau} = O(n^{-k} Z(\log \alpha)) \Leftrightarrow D_\tau = O(k \log_\alpha n - \log_\alpha Z(\log \alpha))$$
(since wlog $D_\tau = O(n)$) $\Leftrightarrow O(n) = O(k \log_\alpha n - \log_\alpha Z(\log \alpha))$.

Now $n!/\alpha^n \leq Z(\log \alpha) \leq n!/\alpha$, so

$$k \log_\alpha n - \log_\alpha Z(\log \alpha) \leq k \log_\alpha n - \log_\alpha (n! \alpha^{-n})$$
$$= k \log_\alpha n - \log_\alpha n! + n \approx k \log_\alpha n - \log_\alpha \left(\left(\frac{n}{e}\right)^n \sqrt{2\pi n}\right) + n$$
$$= k \log_\alpha n - n \log_\alpha n + n \log_\alpha e - \tfrac{1}{2}\log_\alpha 2\pi - \tfrac{1}{2}\log_\alpha n + n$$
$$= O(n \log_\alpha n) \text{ (since wlog } k < n)$$

where we have used Stirling's formula. So the CP becomes $\log_\alpha n = O(1)$; that is, $\alpha = O(n)$. (Note that this requires an energy scale like $\alpha^{O(n)}$ required to perform the annealing.) It is worth noting that $k$ disappears from this estimate: this suggests that relatively small increases in $\alpha$ could dramatically reduce the expected degree for the runtime.

Implicit in all this is the requirement that the rotations induced by the $U$-operators can actually be performed distinguishably: accuracy is key. (It is well known that an exponentially accurate classical computer then can solve NP-problems [Si].) If we write $U \equiv R_\theta$ then $\theta = \cos^{-1}(\alpha^{d/2})$, and the distinguishability requirement can be written as

$$\min \Delta\theta = \frac{\pi}{2}\frac{1}{2^m} = \left.\frac{\partial\theta}{\partial d}\right|_{\alpha=const} \min \Delta d$$

$$\Rightarrow \frac{\log\alpha}{\sqrt{\alpha^d - 1}} \min \Delta d = \frac{\pi}{2^m} \Rightarrow 2^m = \frac{\pi\sqrt{\alpha^d - 1}}{\min \Delta d \log \alpha}$$

$$\Rightarrow m = \log_2\left(\frac{\pi\sqrt{\alpha^d - 1}}{\min \Delta d \log \alpha}\right) = O\left(\log\left(\frac{\sqrt{n^d - 1}}{\min \Delta d \log n}\right)\right).$$

This diverges as $d \to 0$ (which is in some sense not unexpected, as such a situation corresponds to near-degeneracy in the TSP), but is otherwise well behaved; moreover the divergence itself is manifested only on small scales. Hence although the general solution of the TSP as presented will be infeasible (without augmenting the annealing scheme to deal with imperfect phase gating), specific instances of it will generally be feasible to solve on a single quantum computer. Amplitude amplification and general iterative nested search techniques [Ce1] can also be brought to bear on the problem to increase efficiency: by (say) performing repeated quantum searches to return states with likelihoods above given thresholds, and measuring after the last search. In general this can multiply the (polynomial degree of the) complexity class by a small number $a < 1$ (in the limit $a \to 0$ we would approach a logarithmic improvement). This technique can further decrease the runtime for the TSP. Additionally, we can use an auxiliary [classical] buffer to compute the actual distance for a proposed solution and compare it with the previous best result to ensure monotone convergence.

Moreover, initially degenerate instances of the TSP should usually (if not always) also be soluble by decomposing them into subproblems and applying these techniques to appropriate subsets of vertices. By clustering close cities and solving subinstances of the TSP, it is possible to overcome this nondegeneracy criterion to any degree desired without much more computational overhead.

The quantitative degree of improvement this framework offers over "everyday" simulated annealing is unclear. (Chief among the reasons for this [and one which we neglect here] is the energy tradeoff that must be made.) However several factors indicate that it is significant. For example, traditional simulated annealing relies on the Metropolis acceptance criterion [Me], which is in practice problematic. For instance it is very difficult to ensure a significant probability of ending in a global minimum: the inverse temperature must decrease logarithmically with time for this to happen. [GG] As an illustrative example we consider conformational analysis for protein folding: this is an essentially combinatorial technique, and conformational simulated annealing investigations of protein folding take a very long time. The reason for this is that the energy spectrum of a protein is such that the ground state is far below a quasi-continuum of higher-energy states. [Le2, SSK] Classical simulated annealing therefore has limited applicability to this problem, since it cannot access the entire energy spectrum at once but is restricted to local perturbations. (Generalized simulated annealing is also problematic, since although it is not restricted to local perturbations it cannot access the entire space of candidate solutions.) On the other hand, it is quite possible that quantum simulated annealing along the lines drawn above would prove to be an efficient technique for attacking the protein folding problem. The dynamical nature of protein folding (long-duration intermediate regimes are observed in Monte Carlo simulations) also suggests that nested searching could further improve quantum algorithms in this context. [SSK]

It also seems possible to generalize quantum simulated annealing to quantum variants of generalized simulated annealing (GSA), where instead of a Gibbs distribution another probability distribution is used. [AS], [Pe] It has recently been shown that for highly complex systems GSA performs better than the Kirkpatrick-Metropolis incarnation of simulated annealing [XG], and it has been known for some time that GSA offers improved performance for the TSP. [Pe] QGSA requires $\phi$, $F$ satisfying a relationship of the form

$$\phi = \frac{qF(\prod q)}{\sum q}; \quad \sum \phi = F.$$

It is worth noting that the main portion of this analysis does not depend on a specific encoding scheme for paths, and that if a computationally *and* energetically efficient method for realizing the partition function of a problem exists, then quantum simulated annealing will give exact solutions with high probability in expected polytime. Nevertheless, the time-energy tradeoff, first observed by [Ce2], seems to be a Heisenberg-type restriction, and its apparent naturalness gives heuristic grounds to doubt the possibility of realizing an efficient method for realizing combinatorially complex partition functions. This doubt is also corroborated by the results of [LB] for Ising model simulation.

The (admittedly unlikely) prospect of coupling a more energetically efficient simulated annealing sorting algorithm (based on identifying transpositions with spins in the Ising model on a triangular lattice, and which could be used to obtain a *nearly* sorted list very quickly) with a quantum version of the bubblesort (or any other algorithm which is very efficient for nearly sorted lists) is interesting: the ubiquity of sorting as a subroutine suggests that specialized chips with a physical representation of the algorithm as qubits in a triangular configuration and a register to record spin flips could (at least on a small scale) be a common component in some real quantum processors. More generally physical approaches to quantum algorithms (e.g., [FG]) hold promise and are deserving of future study.

The author wishes to thank J. Heagy and D. Ford for many helpful discussions, and C. Bennett, S. Carlson, P. Hermann and P. Anspach for pointing out several crucial issues pertaining to probabilistic quantum algorithms and simulated annealing. This work was performed under IDA Quantum Computing CRP 2048.

[AL]   Abrams, D. S. and Lloyd, S. *Phys. Rev. Lett.* **79**, 2586 (1997).
[AS]   Andricioaei, I. and Straub, J. E. *Phys. Rev. E* **53**, 3055 (1996).
[BBB]  Bennett, C. H., Bernstein, E., Brassard, G. and Vazirani, U. quant-ph/9701001 (1997).
[BV]   Bernstein, E. and Vazirani, U. in *Proc. 25th Ann. ACM Symposium on Theory of Computing.* ACM, New York (1993).
[Ce1]  Cerf, N. J., Grover, L. K. and Williams, C. P. *Phys. Rev. A* **61**, 032303 (2000).
[Ce2]  Cerny, V. *Phys. Rev. Lett.* **48**, 116 (1993).
[FG]   Farhi, E. and Goldstone, J. quant-ph/0007071 (2000).
[GG]   Geman, S. and Geman, D. IEEE Trans. Pattern. Anal. Mach. Intell. **PAMI-6**, 721 (1984).
[HP]   Hogg, T. and Portnov, D. quant-ph/0006090 (2000).
[Hu]   Huntsman, S., unpublished (2000).
[KS]   Kirkpatrick, S. and Selman, B. *Science* **264**, 1297 (1994).
[Ki]   Kitaev, A. Y. quant-ph/9511026 (1995).
[LB]   Lidar, D. A. and Biham, O. *Phys. Rev. E* **56**, 3661 (1997).
[Le1]  Leach, A. *Molecular Modeling—Principles and Applications.* Addison-Wesley Longman, Reading, Massachusetts, 1996.
[Le2]  Levinthal, C. *J. Chem. Phys.* **65**, 44 (1968).
[Me]   Metropolis, N. M., *et al. J. Chem. Phys.* **21**, 6 (1953).
[Pe]   Penna, T. J. P. *Phys. Rev. E* **51**, 1 (1995).
[Ru]   Rudolph, T. quant-ph/9603001 (1996).
[Si]   Simon, J. in *Proc. of the 9th Ann. Symposium of the ACM on Theory of Computing.* ACM, New York (1977).
[SSK]  Sali, A., Shakhnovich, E. and Karplus, M. *Nature* **369**, 248 (1994).
[Ts]   Tsallis, C. *J. Stat. Phys.* **52**, 479 (1988).
[XG]   Xiang, Y. and Gong, X. G. *Phys. Rev. E* **62**, 4473 (2000).